\documentclass[twocolumn,aps,showpacs,prb,amsmath,amssymb,floatfix]{revtex4-1}
\usepackage{color}
\usepackage{graphicx}
\usepackage{dcolumn}
\usepackage{bm}
\usepackage{array}
\usepackage{float}
\usepackage{longtable}
\usepackage{mathrsfs}
\usepackage{txfonts}

\newcommand{\mos}{MoS$_2$}
\newcommand{\tise}{TiSe$_2$}

\begin{document}

\title{Superconducting dome in MoS$_2$ and TiSe$_2$ generated by quasiparticle-phonon coupling }

\author{Tanmoy Das}
\affiliation{Graphene Research Center and Department of Physics, National University of Singapore, 2 Science Drive 3, Singapore 117542.}
\author{Kapildeb Dolui}
\affiliation{Graphene Research Center and Department of Physics, National University of Singapore, 2 Science Drive 3, Singapore 117542.}

\date{\today}

\begin{abstract}
We use a first-principles based self-consistent momentum-resolved density fluctuation (MRDF) model to compute the combined effects of electron-electron and electron-phonon interactions to describe the superconducting dome in the correlated MoS$_2$ thin flake and TiSe$_2$. We find that without including the electron-electron interaction, the electron-phonon coupling and the superconducting transition temperature ($T_c$) are overestimated in both these materials. However, once the full angular and dynamical fluctuations of the spin and charge density induced quasiparticle self-energy effects are included, the electron-phonon coupling and Tc
are reduced to the experimental value. With doping, both electronic correlation and electron-phonon coupling grows, and above some doping value, the former becomes so large that it starts to reduce the quasiparticle-phonon coupling constant and Tc, creating a superconducting dome, in agreement with experiments.
\end{abstract}

\pacs{74.20.-z, 74.20.Fg, 74.78.Na,73.22.-f}

\maketitle\narrowtext

\section{Introduction}

Unconventional superconductivity in cuprates, pnictides, and heavy fermions often reaches its optimum value near the quantum critical point (QCP) of a magnetic ground state, providing a perspective that critical phase fluctuations of the intertwined electronic order drive unconventional  superconductivity.\cite{QCP} However, looking back at conventional superconductors, superconducting (SC) dome is not an unfamiliar feature. It has been observed in Li metal under pressure,\cite{Li}, 
doped SrTiO$_3$,\cite{SrTiO3} gated LaAlO$_3$/SrTiO$_3$ (LAO/STO) interface,\cite{LAOSTO} pressure tunned Fe metal,\cite{Fe}
and also in $T$Se$_2$ ($T$=Ti, Ta and Nb) families,\cite{TiSe2_Dome,TiSe2_QCP,NbSe2,TaSe2} and more recently in thin flake of {\mos}.\cite{MoS2_SC,MoS2_Dome} More interestingly, no evidence of an intertwined electronic order in the SC state is reported in these materials, except in $T$Se$_2$.  The possible role of charge density wave (CDW) in the SC dome in the latter family has also recently been called into question by a pressure dependent x-ray scattering measurement,\cite{TiSe2_noQCP} because the  QCP of CDW in 1{\it T}-{\tise} lies at a higher pressure than the termination of the SC dome. Therefore the presence  of a SC dome with and without an intervening QCP leads to the fundamental question: Can there be an alternative and universal origin of the SC dome which is applicable to all families of superconductors? 

\begin{figure}[t]
\rotatebox[origin=c]{0}{\includegraphics[width=.85\columnwidth]{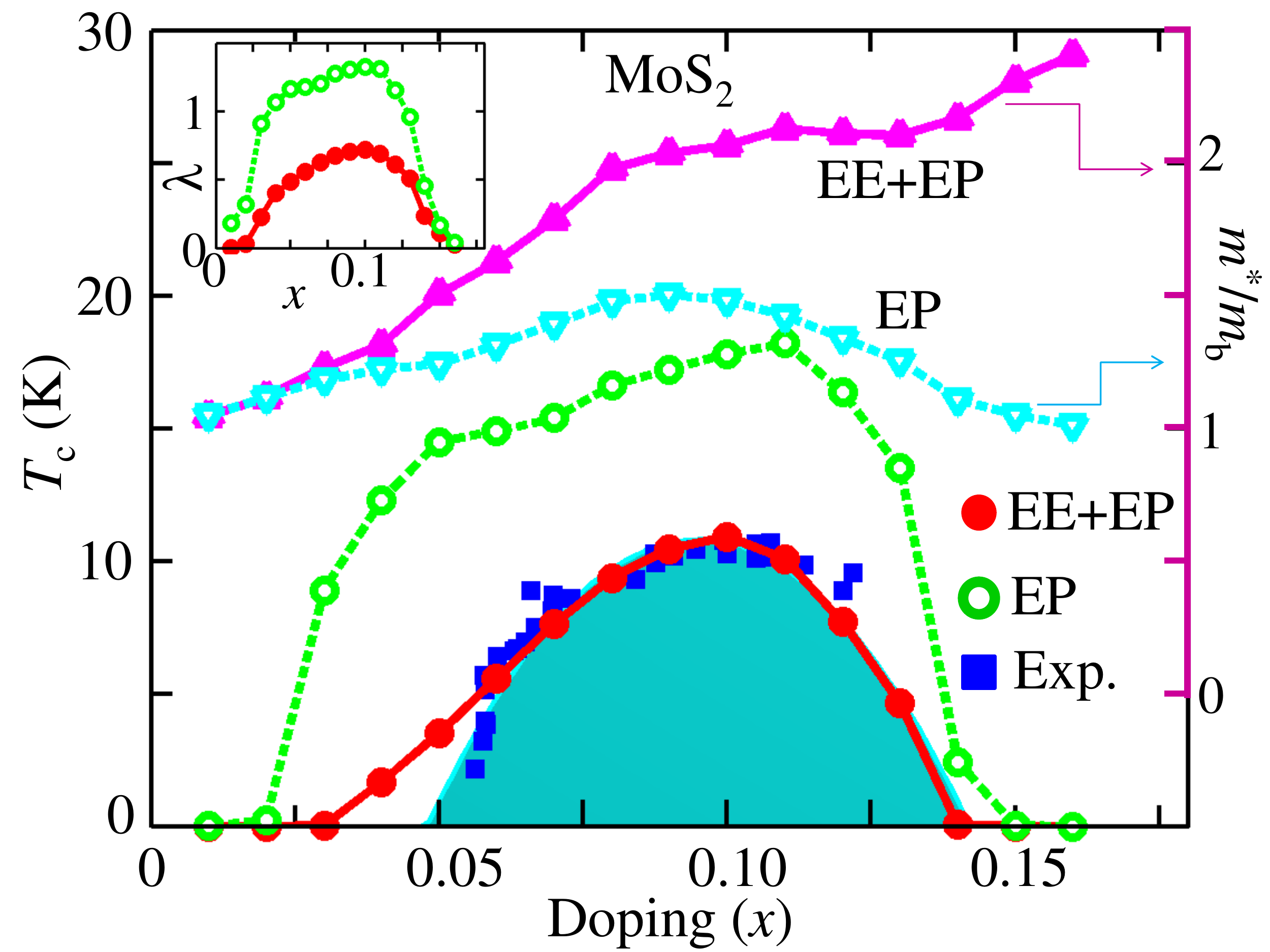}}
\caption{(Color online) SC phase diagram of thin flake of MoS$_2$. Blue squares are experimental $T_c$ data, taken from Ref.~\protect\onlinecite{MoS2_Dome}. Red filled circles give computed $T_c$ by including self-consistent EE and EP interactions. Green open circles give the EP coupling induced $T_c$ when the EE interaction is ignored. Magenta filled triangles and cyan open triangles, respectively, are the computed mass renormalizations for the corresponding two cases. Light blue shading is guide to the eyes to the experimental SC dome. {\it Inset:} EP coupling constant ($\lambda$) when EE interaction is included (red filled circles) and ignored (green open circles).   
}\label{fig1}
\end{figure}

We explore the possible role of the momentum dependent density fluctuations (MRDF) in renormalizing the electronic pairing strength and thereby producing optimized SC transition temperature ($T_c$) as a function of doping. Spin and charge fluctuations are among the two dominant electron-electron (EE) correlations which are ubiquitous in weakly to strongly correlated materials, irrespective of the formation of a static electronic order.\cite{Moriya,AIPDas} These fluctuations renormalize the quasiparticle states, which in turn renormalizes the electron-phonon (EP) or more appropriately quasiparticle-phonon coupling strength and vice versa. In this work, we present a self-consistent theory for the combined EE and EP interaction to calculate the electronic self-energy for the band renormalization, the renormalized EP coupling spectral function $\alpha^2F$, and $T_c$ (see Fig.~1 for these results in {\mos}), starting from the materials specific first-principles band structure. We find that the interplay between the EE and EP coupling has a common role in reproducing the SC dome as a function of doping in both {\mos}\cite{MoS2_Dome} and {\tise}\cite{TiSe2_Dome} samples.

The rest of the paper is organized as follows. In Sec.~\ref{SecII}, we present the methodology which includes the details of DFT calculation in Sec.~\ref{SecIIA}, MRDF method of calculating momentum dependent density-density fluctuation and self-energy in Sec.~\ref{SecIIB}, and the quasiparticle-phonon coupling constant and $T_c$ calculations in Sec~\ref{SecIIC}, and finally the relevant parameters in Sec.~\ref{SecIID}.  The results of band structure, $\alpha^2F$, and $T_c$ are presented in Sec.~\ref{SecIII} for MoS$_2$ thin flake and for bulk TiSe$_2$. Finally, we conclude in Sec.~\ref{SecIV}. Some additional results such as spin and charge fluctuations, momentum dependent self-energies, doping dependent mass renormalizations, $\omega_{\rm log}$, and $\mu^*$ are presented in Appendix.

\section{Method}\label{SecII}

\subsection{Density function theory calculation}\label{SecIIA}

Electronic structure calculations are carried out using density functional theory (DFT) within the local density approximations (LDA) in the Caperley-Alder  parameterization~\cite{LDA} for exchange-correlation functional as implemented in Viena Abinitio Simulation Package (VASP)~\cite{VASP}. Projected augmented-wave (PAW)~\cite{PAW} pseudopotentials are used to describe core electrons. The conjugate gradient method is used to obtain relaxed geometries. Both atomic positions and cell parameters are allowed to relax, until the forces on each atom are less than 1 meV/\AA. The relaxed lattice parameters are: $a=b$ = 3.121 ~\AA, $c/a$=3.868 (bilayer distance) for MoS$_2$, and $a=b$ = 3.437~\AA, $c/a$=1.693 for TiSe$_2$.

The kinetic energy cutoff is fixed at 650 eV. $\Gamma$ centered $k$-mesh grids of 14$\times$14$\times$1$\times$ and 14$\times$14$\times$8$\times$ are used in the self-consistent calculations for MoS$_2$ and TiSe$_2$, respectively.  Note that we did not consider any van der Waals (vdW) correction in our calculations,  as the previous vdW+DFT calculations show that bilayer MoS$_2$ exhibits an indirect gap between the $\Gamma$ to K direction, which contradicts  GW calculations~\cite{vdW, GW}. However our calculated LDA band structure correctly describes this feature. Spin-orbit correction (SOC) is included  in the band structure calculations. We note that for bilayer MoS$_2$, the SOC is significantly reduced due to the presence of the inversion symmetry~\cite{SOC}. Furthermore, in order to study the lattice dynamics, force constants are calculated for 3$\times$3$\times$1 and 3$\times$3$\times$2  super-cell  of MoS$_2$ and TiSe$_2$, respectively within the framework of density functional perturbation theory~\cite{DFPT} using the VASP code. Subsequently, phonon dispersions are calculated using Phonopy package~\cite{phonopy}.

The band structure for monolayer and bilayer MoS$_2$ has significant differences.\cite{MoS2band,MoS2GWbulk}  In monolayer system, both the conduction band bottom and the valence band top lie at the K point. In bilayer system, the bottom of the conduction band shifts to somewhere in between the $\Gamma$ to K points, and the top of the valence band moves to the $\Gamma$ point. However, from bilayer to the bulk system, the band structure does not change significantly.  Therefore, given that the experiment is performed on a thin flake of MoS$_2$ sample, we use the band structure for the bilayer system. 

The band dispersion between LDA and LDA+GW calculation does not change characteristically,\cite{GW,MoS2GWbulk} which means the doping dependence of the density of states would also be similar for both cases at the expense of  a constant energy shift. Therefore, our many body and $T_c$ calculations would yield similar results for the LDA and LDA+GW band structure inputs.

Finally, we have used a rigid band shift for the DFT dispersion, whereas with self-energy correction, the new Fermi level is recalculated self-consistently. For MoS2, the carrier density is monitored experimentally by gating, which is more appropriate to be modeled by rigid-band shift. A previous DFT calculation by one of the present author\cite{kapildeb} showed that even an absorption alkali metal can rigidly shift the Fermi level to the conduction band in the same way. For TiSe2, the SC dome is observed experimentally by Cu intercalation. However, as shown in the same experimental paper (Fig. 5 of Ref.~\onlinecite{TiSe2_Dome}), the Sommerfield coefficient of the same sample changes almost linearly with doping. This result is consistent with our finding of the density of states at the Fermi level with EE+EP correlation [our Fig.~\ref{fig4}(d)]. This supports that rigid band shift method is applicable in this family as well.

\subsection{Momentum resolved density fluctuation (MRDF) theory}\label{SecIIB}

The VASP band structure information is directly implemented within our MRDF code. The single-particle Green's function is defined as ${\tilde G}_0({\bf k},i\omega_\mathfrak{n})=\left(i\omega_{\mathfrak{n}}{\tilde 1}-{\tilde H}\right)^{-1}$, where $i\omega_{\mathfrak{n}}$ is the Matsubara frequency for the fermions, and $H$ is the non-interacting Hamiltonian constructed by downfolding the DFT bands into the low-energy energy levels. The explicit form of $G$ is then obtained as
\begin{equation}\label{G0}
G_{mn}({\bf k},i\omega_{\mathfrak{n}}) = \sum_{\nu}\frac{\phi^{\nu}_{{\bf k},m}\phi^{\nu\dag}_{{\bf k},n}}{i\omega_{\mathfrak{n}}-\xi^{\nu}_{\bf k}}.
\end{equation}
Here ${\bf k}$ and $\omega$ are the quasiparticle momentum and frequency, and ${\bf q}$ and $\omega_p$ are the bosonic excitation momentum and frequency, respectively. $\phi^{\nu}_{{\bf k},m}$ is the eigenstate for the $\nu^{\rm th}$ DFT band ($\xi^{\nu}_{\bf k}$), projected onto the $m^{\rm th}$ orbital. The non-interacting density fluctuation susceptibility in the particle-hole channel represents joint density of states (JDOS), which can be calculated by convoluting the corresponding Green's function over the entire Brillouin zone (BZ) to obtain (spin and charge bare susceptibility are the same in the paramagnetic ground state)\cite{Graser}:
\begin{eqnarray}\label{chibare}
\chi_{0,mn}^{st}({\bf q},\omega_{\mathfrak{p}})&=&-\frac{1}{\Omega_{\rm BZ}\beta}\sum_{{\bf k},\mathfrak{n}}G_{mn}({\bf k},i\omega_{\mathfrak{n}}) \nonumber\\
&&~~~~~~~~~~~~\times G_{st}({\bf k+q},i\omega_{\mathfrak{n}}+\omega_{\mathfrak{p}}),
\end{eqnarray}
where $\beta=1/k_BT$, and $k_B$ is the Boltzmann constant and $T$ is temperature. $\Omega_{\rm BZ}$ is the electronic phase space volume. $f^{\nu}_{\bf k}$ and $n_p$ are the fermion and boson occupation numbers, respectively.  After performing the Matsubara summation over the fermionic frequency $\omega_n$ and taking analytical continuation to the real frequency as $\omega_\mathfrak{n}\rightarrow\omega+i\delta$, we get
\begin{eqnarray}\label{chifull}
\chi_{0,mn}^{st}({\bf q},\omega_{\mathfrak{p}})&=&-\frac{1}{\Omega_{\rm BZ}}\sum_{{\bf k},\nu,\nu^{\prime}}\phi^{\nu\dag}_{{\bf k}+{\bf q},s}\phi^{\nu}_{{\bf k}+{\bf q},t}\phi^{\nu^{\prime}}_{{\bf k},n}\phi^{\nu^{\prime}\dag}_{{\bf k},m} \nonumber\\
&&~~~~~~\times\frac{f^{\nu}_{{\bf k}+{\bf q}} - f^{\nu^{\prime}}_{{\bf k}} }{\omega_{\mathfrak{p}}+i\delta -\xi^{\nu^{\prime}}_{{\bf k}} +\xi^{\nu}_{{\bf k}+{\bf q}}}.
\end{eqnarray}

The interacting density-density correlation functions are computed within the random phase approximation which includes multiband components of the electronic interaction including intra- and inter-orbital Coulomb interactions, $U$ and $V$, as well as Hund's coupling $J_H$, and pair-exchange term $J^{\prime}$:
\begin{widetext}
\begin{eqnarray}
H_{\rm int}&=&\sum_{{\bf k}_1-{\bf k}_4}\left[U\sum_m c^{\dag}_{{\bm k}_1,m\uparrow}c_{{\bf k}_2,m\uparrow}c^{\dag}_{{\bf k}_3,m\downarrow}c_{{\bm k}_4,m\downarrow}
+\sum_{m<n,\sigma}\left(Vc^{\dag}_{{\bf k}_1,m\sigma}c_{{\bf k}_2,m\sigma}c^{\dag}_{{\bf k}_3,n\bar{\sigma}}c_{{\bf k}_4,n\bar{\sigma}}
+(V-J_H)c^{\dag}_{{\bf k}_1,m\sigma}c_{{\bf k}_2,m\sigma}c^{\dag}_{{\bf k}_3,n\sigma}c_{{\bf k}_4,n\sigma}\right)\right.\nonumber\\
&&\left. +\sum_{m<n,\sigma}\left(J_Hc^{\dag}_{{\bf k}_1,m\sigma}c^{\dag}_{{\bf k}_3,n\bar{\sigma}}c_{{\bf k}_2,m\bar{\sigma}}c_{{\bf k}_4,n\sigma}
+ J^{\prime} c^{\dag}_{{\bf k}_1,m\sigma}c^{\dag}_{{\bf k}_3,m\bar{\sigma}}c_{{\bf k}_2,n\bar{\sigma}}c_{{\bf k}_4,n\sigma} + h.c.\right)\right].
\label{intH}
\end{eqnarray}
\end{widetext}
Here $c^{\dag}_{{\bf k}_1,m\sigma}$ ($c_{{\bf k}_1,m\sigma}$) is the creation (annihilation) operator for an orbital $m$ at crystal momentum ${\bf k}_1$ with spin $\sigma=\uparrow$ or $\downarrow$, where $\bar{\sigma}$ corresponds to opposite spin of $\sigma$. In the multiorbital spinor, the above interacting Hamiltonian can be collected in a interaction tensor $\tilde{U}_{s/c}$, where the subscripts $s$, $c$ stand  spin and charge density fluctuations. The nonzero components of the matrices $\tilde{U}_c$ and $\tilde{U}_s$ are given as\cite{Graser,Ueda}
\begin{eqnarray}\label{RPAU}
&&\tilde{U}^{mm}_{s,mm}= U,~~~\tilde{U}^{mm}_{s,nn}= \frac{1}{2}J_H,\nonumber\\
&&\tilde{U}^{mn}_{s,mn} = \frac{1}{4}J_H+V, ~~~\tilde{U}^{nm}_{s,mn}=J^{\prime},\nonumber\\
&&\tilde{U}^{mm}_{c,mm}= U,~~~\tilde{U}^{mm}_{c,nn}= 2V,\nonumber\\
&&\tilde{U}^{mn}_{c,mn}
= \frac{3}{4}J_H-V, ~~~\tilde{U}^{nm}_{c,mn}=J^{\prime}.\nonumber\\
\end{eqnarray}
Of course, it is implicit that all the interaction parameters are orbital dependent. Within the RPA, spin and charge channels become decoupled. The collective many-body corrections of the density-fluctuation spectrum can be written in matrix representation: $\tilde{\chi}_{s/c}=\tilde{\chi}^0[\tilde{1}\mp\tilde{U}_{s/c}\tilde{\chi}^0]^{-1}$, for spin and charge densities, respectively. 
$\tilde{\chi}^0$ matrix consists of components  $\chi_{0,mn}^{st}$ with the same basis in which the interactions $\tilde{U}_{s/c}$ are defined above. 

The interaction parameters ($U$, $V$, $J_H$, and $J^{\prime}$) are not parameterized individually. Rather, we estimate the components of the interaction matrices $\tilde{U}_{s/c}$ within the Kanamori criterion $\tilde{U}_s\le \big({\rm max}[ \tilde{\chi}^{\prime}_{0}({\bf q},0)]\big)^{-1}$, and $\tilde{U}_s=\tilde{U}_c$, and set the values at the optimal doping for each system. Note that in the self-consistent loop when the self-energy is included in $\tilde{\chi}^{\prime}_{0}$, the corresponding values of $U$s effectively include the screening phenomena due to spin, charge and phonon scatterings. 

Finally, the EE interaction potentials for the electronic state are computed as 
\begin{equation}
V_{mn,i}^{st}({\bf q},\omega_{\mathfrak{p}}) = \frac{\eta_{i}}{2}\left[\tilde{U}_i\tilde{\chi}_{i}^{\prime\prime}({\bf q},\omega_{\mathfrak{p}})\tilde{U}_i \right]_{mn}^{st},
\end{equation}
 where $i$ stands for spin and charge components, $\eta$ = 3, 1 for the spin and charge channels, respectively. The electron-phonon coupling effect is calculated similarly:
\begin{equation}\label{EP}
V_{mn,p}^{st}({\bf q},\omega_{\mathfrak{p}})=\sum_{{\bf k},\mu}\left|g_{ms}^{\mu}({\bf k},{\bf q})\right|\left|g_{nt}^{\mu}({\bf k},{\bf q})\right|\delta(\omega_{\mathfrak{p}}-\omega_{q\mu}),
\end{equation}
where $\omega_{q\mu}$ is the phonon dispersion for band $\mu$, and the subscript `p' stands for the EP term. The EP coupling matrix-element is $g_{ij}^{\mu}({\bf k},{\bf q}) = \sum_{\nu}  g^{\mu}_{{\bf q},0}\phi_{{\bf k},i}^{\nu\dag}\phi_{{\bf k}+{\bf q},i}^{\nu}$, where the momentum averaged EP scattering amplitude $g_0$ is deduced from the first-principles calculation.

The feedback effect of the two EE potentials, and the EP coupling on the electronic spectrum is then calculated via self-energy calculation within the MRDF method\cite{EPselfenergy,AIPDas,PuDas}
\begin{eqnarray}
\label{SE}
\Sigma_{mn,i}({\bf k},\omega) = \frac{1}{\Omega_{\rm BZ}}\sum_{{\bf q},st,\nu}\int_{-\infty}^{\infty} d\omega_p V_{mn,i}^{st}({\bf q},\omega_p)\Gamma_{mn,\nu}^{st}({\bf k},{\bf q})\nonumber\\
\times \left[\frac{1-f^{\nu}_{{\bf k}-{\bf q}} +n_p}{\omega+i\delta -\xi^{\nu}_{{\bf k}-{\bf q}} -\omega_p}
+\frac{f^{\nu}_{{\bf k}-{\bf q}} +n_p}{\omega+i\delta -\xi^{\nu}_{{\bf k}-{\bf q}} +\omega_p} \right],
\end{eqnarray} 
where the subscript $i$ stands for spin, charge and phonon contributions. The vertex correction $\Gamma_{mn,\nu}^{st}({\bf k},{\bf q})$ encodes both the angular and dynamical parts of the vertex, which are combined to obtain $\Gamma_{mn,\nu}^{st}({\bf k},{\bf q})=\phi^{\nu\dag}_{{\bf k}-{\bf q},s}\phi^{\nu}_{{\bf k}-{\bf q},t}(1-\partial \Sigma_{mn}({\bf k}-{\bf q},\omega)/\partial \omega)_{0}$. Full self-consistency requires the bare Green's function $G_0$ in Eq.~(\ref{G0}) to be replaced with the self-energy dressed ${\tilde G}^{-1}({\bf k},\omega)={\tilde G}_0^{-1}({\bf k},\omega)-{\tilde \Sigma}({\bf k},\omega)$, where the total self-energy tensor is $\tilde{\Sigma}({\bf k},\omega)=\tilde{\Sigma}_s({\bf k},\omega)+\tilde{\Sigma}_c({\bf k},\omega)+\tilde{\Sigma}_p({\bf k},\omega)$,y and calculate susceptibilities and self-energies with the dressed Green's function until the self-energies converges. This procedure is numerically expensive, especially in the multiband systems and when full momentum dependence is retained. Therefore, we adopt a modified self-consistency scheme, where we expand the real part of the total self-energy tensor as $\tilde{\Sigma}^{\prime}({\bf k},\omega) = (1-\tilde{Z}_{\bf k})^{-1}\omega$ in the low-energy region [$|\omega|<0.2 - 0.3$~eV in the present materials]. The resulting self-energy dressed quasiparticle dispersions ${\bar \xi}^{\nu}({\bf k}) = Z^{\nu}_{\bf k}\xi^{\nu}({\bf k})$ are used in Eqs. (\ref{G0})-(\ref{SE}), which keep all the formalism unchanged with respect to the momentum resolved orbital selective quasiparticle renormalization factor ${\tilde Z}_{\bf k}$.

\subsection{Quasiparticle-phonon coupling and $T_c$ calculation}\label{SecIIC}

In the systems where the interaction strength is of the order of the bandwidth, i.e., the kinetic and potential energies are of the same order, various instabilities develop. Different materials having different lattice structure and Fermi surface topology are prone to different forms of instabilities, among which leading contributions usually arise from the superconductivity and density wave fluctuations in the spin, charge and lattice sectors. These instabilities often lead to an incoherent, or a gapped Fermi surface at the expense of superconductivity and / or a static density wave (s).

Therefore, even for conventional superconductors, especially for the $d$-electronic systems, such correlated electronic structure or quasiparticle spectrum can lead to substantial modification of the EP coupling constant, as originally proposed by Eliashberg.\cite{Eliashberg} The same effect also leads to a violation of the Migdal's theorem.\cite{Migdalbreakdown} We treat this problem using a quasi-perturbation method in which the non-interacting electron and phonon dispersions are computed via DFT framework. Then the EP coupling vertex is calculated self-consistently in which the electronic state is dressed with both EE (density fluctuations) and EP coupling. The quasiparticle-phonon coupling density of states (DOS) or the so-called $\alpha^2F$ is calculated as\cite{Eliasberg}
\begin{equation} \label{alp2Fp}
\alpha^2F_p(\omega_p) = \frac{1}{\Omega_{BZ}{\bar N}(0)}\sum_{{\bf q},{\bf k},\nu}{\rm Tr}\big[\tilde{V}_p({\bf q},\omega_p)\big]\delta({\bar \xi}_{\bf k}^{\nu})\delta({\bar \xi}_{{\bf k}+{\bf q}}^{\nu}),
\end{equation}
where ${\bar N}(0)$ is the correlated electronic DOS at the Fermi level, and  the self-energy dressed quasiparticle band is ${\bar \xi}_{\bf k}^{\nu}={\xi}_{\bf k}^{\nu}+\Sigma_{\nu}^{\prime}({\bar{\xi}_{\bf k}^{\nu}})$, and $V_p$ is defined in Eq. (\ref{EP}) but with self-energy correction. Therefore, the SC transition temperature becomes\cite{AllenDynes}
\begin{eqnarray} \label{Tc}
T_c &=& \frac{\omega_{\rm log}}{1.2k_B}\exp{\left(-\frac{1.04(1+\lambda)}{\lambda-\mu^*(1+0.62\lambda)}\right)},
\end{eqnarray}
where the traditional Debey frequency is replaced by\cite{Dynes}
\begin{equation}\label{wlog}
\omega_{\rm log}=\exp{\left(\frac{2}{\lambda}\int d\omega_p\log{\omega_p}\frac{\alpha^2F_p(\omega_p)}{\omega_p}\right)},
\end{equation}
and the EP coupling strength for the Cooper pair is
\begin{equation}\label{lambda}
\lambda = 2\int d\omega_p \frac{\alpha^2F_p(\omega_p)}{\omega_p}.
\end{equation}
Note that the EP coupling in the SC channel, $\lambda$, is different from the EP coupling constant for the single particle spectrum $\lambda_{qp}=Z^{-1}-1$, where $Z^{-1}=1-(\partial\Sigma^{\prime}/\partial\omega)_{\omega=0}$, that one often obtains from the quasiparicle self-energy.\cite{Bob_opticalglue}

The Coulomb pseudopotential $\mu^*_0$ is often taken as a parameter, however, the evolution of it in correlated system and as a function of doping can be estimated as
\begin{equation}\label{mu*}
\mu^*=\frac{\mu^*_0}{1+\mu^*_0{\rm  \log}\left(\frac{\bar{U}}{\omega_{\rm log}}\right)},
\end{equation}
 where the renormalized Coulomb interaction is $\bar{U}=U/(1+\lambda)$. The doping dependent values of $\omega_{\rm log}$ and $\mu^*$ are shown in Fig.~\ref{fig8} for both materials.

\subsection{Parameters}\label{SecIID}
The band structure and the phonon spectrum are deduced from first-principles calculations without any adjustable parameters. All other parameters are adjusted only at the optimal doping for the corresponding systems to get the experimental value of $T_c$. For example, we deduce the Coulomb interactions from the Kanamori criterion from the self-energy dressed `bare' susceptibilities (here `bare' refers to the susceptibility before invoking RPA effect, but with including self-energy). The values are set at the optimal doping, and are kept to be doping independent, but material dependent. The largest interaction value we find is 2~eV for both cases.  The EP coupling potential $g_0$ is found to be 33 meV for MoS$_2$ and for 35 meV TiSe$_2$ at their optimal dopings. Finally, for the bare value of $\mu^*_0$ in Eq.~(\ref{mu*}), we use a standard value $\mu^*_0=0.1$ for both materials. All the other values, such as renormalized $\omega_{\rm log}$, $\mu^*$, and $\bar{U}$ are computed self-consistently at all dopings using the formulas given above.

\section{ Results}\label{SecIII} 

\subsection{MoS$_2$ thin flake}\label{SecIIIA}

\begin{figure}[here]
\rotatebox[origin=c]{0}{\includegraphics[width=0.85\columnwidth]{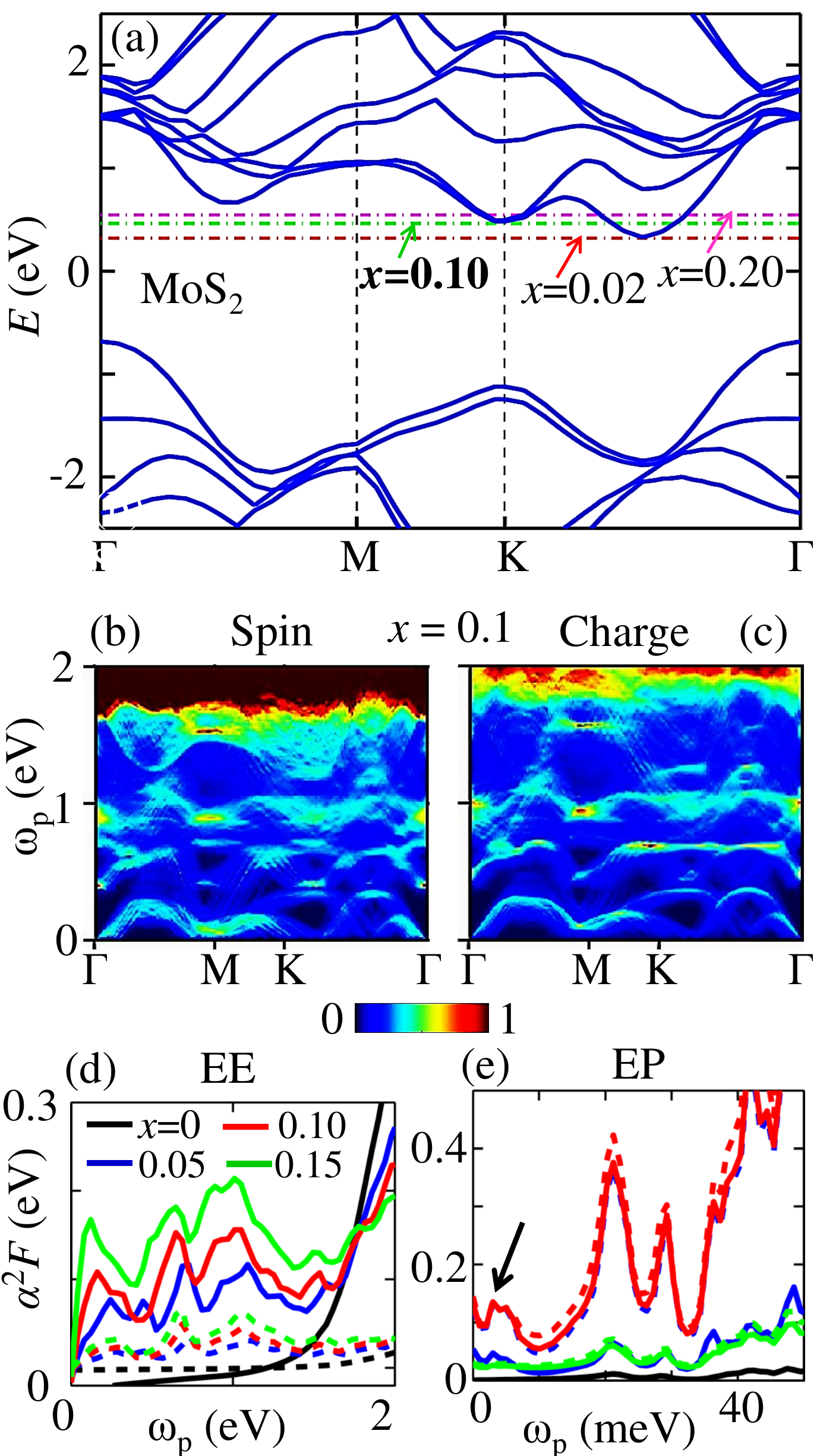}}
\caption{(Color online) (a) DFT band structure of bilayer MoS$_2$. (b)-(c) Imaginary part of the spin- and charge-fluctuations ($\chi^{\prime\prime}_{s/c}$) dispersion along the high-symmetry directions ($q$), respectively at optimal doping $x$=0.1. (d) Self-consistently evaluated electronic $\alpha^2 F(\omega)$ is plotted for various representative dopings. Solid lines are the total value (spin + charge), while the dashed lines are their charge contributions. (e) Doping evolution of the EP coupling $\alpha^2F$ in the presence of self-consistent EE and EP interactions. Dashed lines are the same self-consistent EP $\alpha^2F$ but without EE interaction. Arrow dictates the presence of multiple acoustic phonon modes with finite doping.
}\label{fig2}
\end{figure}

For {\mos}, SC is observed in a thin flake sample. As mentioned in Sec.~\ref{SecIIA} above, to simulate thin flake sample, we use band structure for the bilayer MoS$_2$ (see Fig.~\ref{fig2}(a)). In the band insulator {\mos} at $x$=0, susceptibility is fully gapped below the particle-hole continuum [Figs.~\ref{fig5}(a-b)], and thus no significant renormalization arises from the EE part. At finite doping, as shown for optimal doping $x$=0.1 in Figs.~\ref{fig2}(b-c), several new dynamical excitation channels arise at low-energy, mainly dominated by the intraband transitions across $E_F$. The spin channel moves to lower energy than the charge one, and possess larger intensity, giving largest contribution to the many-body renormalization effect.

We show the doping evolution of all three correlation functions by computing the {\bf k}- and orbital averaged correlation spectrum, in the same spirit as Eq.~(\ref{alp2Fp}), for spin and change fluctuations\cite{Bob_opticalglue} as $\alpha^2F_i(\omega_p)=1/\Omega_{\rm BZ}\sum_{{\bf q}}{\rm Tr}\big[\tilde{V}_i({\bf q},\omega_p)\big]$. In Fig.~\ref{fig2}(d), the total electronic (spin + charge) $\alpha^2F$ is plotted for several dopings. We find several dominant peaks at all dopings, except at $x=0$, whose strength increases with doping, suggesting that the strength of the EE correlation gradually increases. On the other hand, the EP $\alpha^2F$, shown in Fig.~\ref{fig2}(e), shows maximum intensity at $x$ = 0.1. Interestingly, the EP $\alpha^2F$ without including EE interaction in the electronic spectrum (dashd lines) shows fairly doping independent behavior from underdoped to the optimally doped region. The resulting EP coupling constant, $\lambda$, exhibits a dome feature with doping as shown in the {\it inset} to Fig.~\ref{fig1}. This result suggests that although superconductivity in MoS$_2$ is unlikely to have electronic mechanism, however, the EE correlation is important in renormalizing the EP coupling which in turn governs the SC dome.

\begin{figure}[top]
\rotatebox[origin=c]{0}{\includegraphics[width=.9\columnwidth]{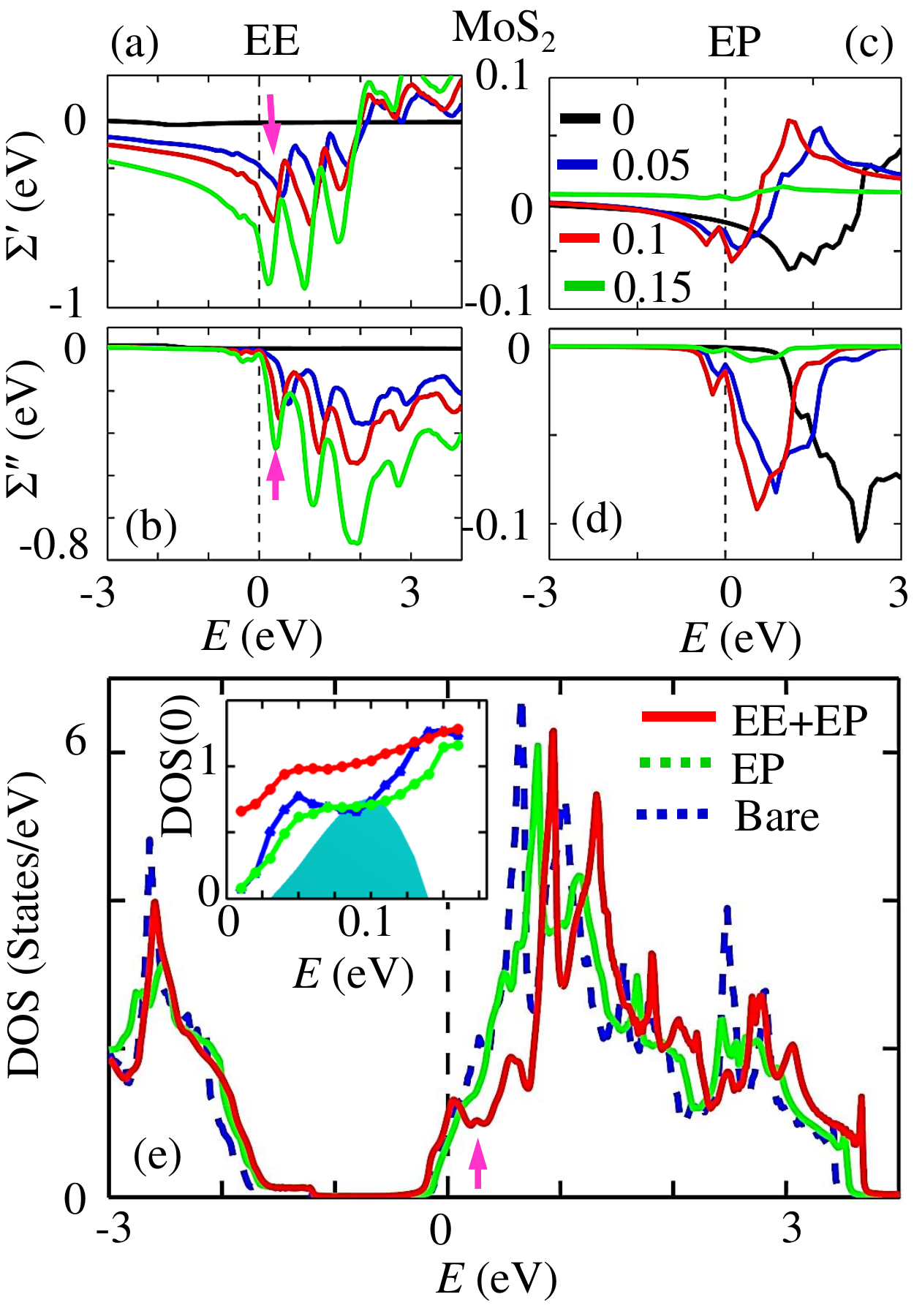}}
\caption{(Color online) (a)-(b) Real ($\Sigma^{\prime}$) and imaginary ($\Sigma^{\prime\prime}$) parts of the EE self-energy at four representative dopings [see legend in (c)]. (c)-(d) Same results, but for EP coupling. (e) DOS at optimal doping. The vertical arrow in (a), (b) and (e) points out that the peak in $\Sigma^{\prime\prime}$ in (d) corresponds to the dip in the DOS in (e). {\it Inset:} The doping evolution of the DOS at $E_F$. Blue symbols are the bare DFT value, while green open circles are DOS with only EP coupling and the red filled circles are the same with including both EE and EP interactions. Light blue shading is the theoretical $T_c$ in arbitrary unit.
}\label{fig3}
\end{figure}

In Fig.~\ref{fig3}, we present the electronic self-energy and the dressed DOS for MoS$_2$. The self-energy is presented for the lowest energy band which crosses $E_F$, and possess largest renormalization. The corresponding ($k$-averaged) mass renormalization factor $m^*/m_b=(1-\partial \Sigma^{\prime}(\omega)/\partial \omega)^{-1}_0$ (where $m_b$ is the DFT band mass) is shown in Fig.~\ref{fig1}. As expected, the electronic $\Sigma$ is very weak at $x$=0. At finite doping, the intra-band excitations provide large mass enhancements, reaching a value above 2 at optimal to overdoped region. The EP contribution, shown in Figs.~\ref{fig3}(c-d) remains comparatively weak throughout the phase diagram, however, reaches a maximum value near the optimal doping. Both EE and EP self-energies show strong particle-hole asymmetry due to the semimetallic-like band structure of this system. The imaginary part of the self-energy, $\Sigma^{\prime\prime}$, possess multiple low-energy peaks, coming from several collective modes in the susceptibility discussed in Fig.~\ref{fig2}. Due to causality, the real part of self-energy, $\Sigma^{\prime}$ sharply changes slope at energies where $\Sigma^{\prime\prime}$ has peaks. 

The self-energy splits the non-interacting DOS into a low-energy quasiparticle peak, and a higher-energy incoherent satellite (or hump) feature as seen in the $\Sigma$-dressed DOS in Fig.~\ref{fig3}(e). This unique self-energy behavior, creating a peak-dip-hump feature, is also observed in cuprates,\cite{AIPDas} and actinides superconductors.\cite{PuDas} Although, EP self-energy possess similar energy dependence, however, it fails to create any accountable peak-dip-hump feature in this system due to its weak strength. The spectral weight transfer from the higher energy states to the quasiparticle states at $E_F$, dictated by the strength of the self-energy effects, can be deduced by the change in DOS between the interacting case (red filled symbol) and non-interacting DOS (blue open circles). This doping dependence of the EE interaction plays a dominant role in creating the dome-like doping dependence of $T_c$, shown in Fig.~\ref{fig1}. 

\begin{figure*}[th]
\rotatebox[origin=c]{0}{\includegraphics[width=2\columnwidth]{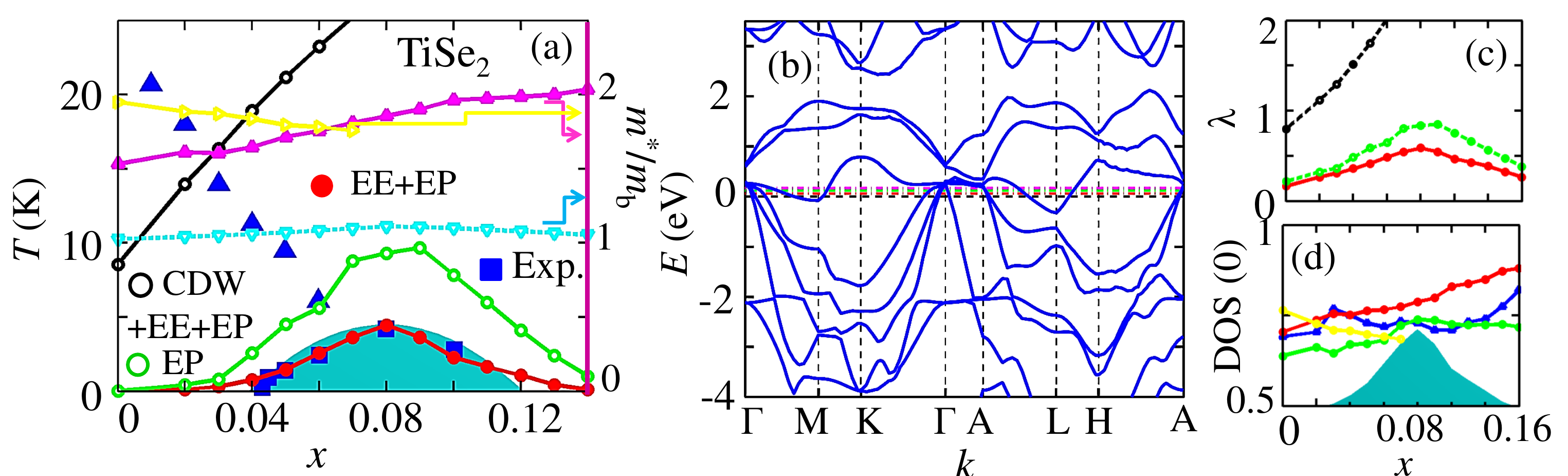}}
\caption{(Color online) (a) SC phase diagram of {\tise}. Blue squares are $T_c$, where blue triangles are $T_{\rm CDW}$ ($\times$10) from Ref.~\protect\onlinecite{TiSe2_Dome}. Red filled and green open circles give computed $T_c$ for the cases of EE+ EP, and EP only respectively, but without CDW. Black open circles give $T_c$ when CDW and EE+EP interactions are all self-consistently included. Magenta filled, cyan open and yellow open triangles give the computed mass enhancement for the three cases discussed above. Light blue shading is guide to eyes to the experimental SC dome behavior. (b) DFT band structure in the 3D momentum space (along high-symmetry directions). Different horizontal dashed lines give $E_F$ for $x$=0, 0.04, 0.08 and 0.12 with increasing values. (c) EP coupling constant for the three cases as discussed in (a) with same symbols. (d) DOS at $E_F$ for bare (blue open triangles), with only EP interaction (green open circles), with EE and EP interactions (red filled circles) and with CDW ($\times$ 4 yellow open circles). Light blue shading is the theoretical $T_c$ in arbitrary unit. 
}\label{fig4}
\end{figure*}

We estimate the value of $T_c$ from the renormalized EP $\alpha^2F$ by using the standard Allen-Dynes formula given Eq.~(\ref{Tc}) above. The so-called Debye frequency, which is modified to a renormalized value as $\omega_{\rm log}$, does not show any significant doping dependence [Figs.~\ref{fig8}(c-d)].\cite{Note_Haas} Similarly, the renormalized screened Coulomb potential $\mu^*$ also does not possess any significant doping dependence for the constant value of $\mu^*_0$ and onsite Coulomb potential $U$.  Using this formula, we find that theoretical estimation of $T_c$ quantitatively reproduces the experimental dome\cite{MoS2_Dome} when both EE and EP interactions are self-consistently included in the renormalization term. When EE interaction is neglected, but the EP self-energy is included in the renormalized $\alpha^2F$ spectrum, the calculation overestimates the experimental $T_c$ in the entire phase diagram (also see calculations of EP coupling induced $T_c$ in {\mos} without including its renormalization in Refs.~\onlinecite{MoS2_EP,Haas}). 

\subsection{TiSe$_2$}\label{SecIIIB}
 
Next we study the origin of SC dome in {\tise}. In Fig.~\ref{fig4}(a) we present our results of $T_c$, calculated for the case of EP, EE + EP, and with CDW and EE + EP interactions. We include CDW at nesting ${\bf Q}$=(1/2,1/2,1/2) as demonstrated experimentally,\cite{TiSe2_noQCP} and the mean-field CDW gap is taken to be $\Delta=13.4k_BT_{\rm CDW}$ at various dopings, which reasonably reproduces the experimental gap.\cite{TiSe2_CDWgap} The self-energy properties and the peak-dip-hump feature in the DOS are all characteristically analogous to {\mos}. Moreover, in {\tise} the spin-component dominates the fluctuation strength, and then follows the contributions of charge and phonon terms in order. 

{\tise} is a three-dimensional system with considerable dispersion along the $k_z$ direction as shown in Fig.~\ref{fig4}(b). Therefore, the correlation effect is reduced here compared to {\mos}, as evident in the lower mass renormalization seen in Fig.~\ref{fig4}(a). As a result, the values of EP coupling constant and $T_c$ with and without including EE interaction are closer to each other in this system. However, the presence of CDW enhances the EP coupling constant above $\lambda>2$ at the optimal doping, as shown in Fig.~\ref{fig4}(c), in which the Migdal theorem is violated.\cite{Migdal} We find that CDW overestimates $T_c$ by about 10-20 times. On the basis of this reasoning, we conclude that CDW and SC are competing in TiSe$_2$. Our finding is consistent with the conclusions of the x-ray scattering data in 1$T$-TiSe$_2$ systems\cite{TiSe2_noQCP}, and also with a DFT calculation.\cite{TiSe2_CDW} Moreover, in QCP induced unconventional superconductors, a linear-in-$T$ dependence of resistivity is observed at the SC dome, which is attributed to non-Fermi liquid behavior. But in TiSe$_2$, the resistivity data continues to exhibit quadratic $T$ dependence throughout the entire SC dome,\cite{TiSe2_Dome,TiSe2_QCP,NbSe2,TaSe2} pointing against the existence of a QCP. Therefore the quasiparticle renormalization and the spectral weight transfer between the quasiparticle state and the high-energy states, driven by spin- and charge fluctuations, play a dominant role in reducing $T_c$ in conventional superconductors.

With the same interaction parameters which give good estimation of $T_c$, we also get mass renormalization in good accord with experiments in both systems. For MoS$_2$, we find $m^*/m_b\sim$ 2 - 2.5 in the optimal to overdoped region. The corresponding experimental value, deduced by ARPES\cite{ARPESMoS2} with respect to a simple parabolic band, is $\sim$2.4 for the hole band. Again for TiSe$_2$, we estimate the experimental mass renormalization by comparing ARPES data\cite{ARPESTiSe2} with the DFT band, and find an average value for two low-lying bands is $m^*/m_b\sim$1.74, which is close to our result of 1.8 - 2. It is interesting to notice that despite having similar mass renormalizations, $T_c$ values differ substantially in two systems. This is because EE interaction does not solely determine the value of $T_c$, but it reduces the EP coupling strength. For MoS$_2$ at optimal doping, EP coupling constant without EE interaction is $\sim$1.3 which reduces  by a factor of $\sim$0.53 with when EE interaction is included, and the mass enhancement at the same doping is $\sim$2. For TiSe$_2$ at optimal doping, these corresponding numbers are 0.85, 0.65 and 1.8, respectively.

\section{Conclusion}\label{SecIV}
The main conclusion of this work is that the EP coupling constant is substantially modified by the dynamical EE interactions, which are, in general, tunable via external parameters such as doping, pressure, magnetic field, and others. While the examples are demonstrated here for two families of conventional superconductors, generalization to other forms of electron-boson coupling, such as spin-fluctuation or polaron mediated superconductivity is rather straightforward. Doping a Mott insulator usually weakens the EE correlation strength as seen in cuprates.\cite{AIPDas} On the other hand, as demonstrated here, doping enhances EE interactions in the band insulator. In both Mott and band insulators, superconductivity seemingly emerges when the EE interaction strength falls into the intermediate coupling region where sufficient spectral weight is transfered to the quasiparticle states near $E_F$ from the higher energy `incoherent' hump (s) via coupling to dominant spin and charge fluctuations.\cite{AIPDas,ASWT} We find that when these EE interactions are taken into account in the EP (or, more appropriately, quasiparticle-phonon) coupling, its strength and $T_c$ are significantly changed and acquires doping dependence. Taken together, our study suggests that as a method of controlling the quasiparticle-boson couping to optimize $T_c$, EE interactions may play the common role in creating the SC dome behavior in both conventional and unconventional superconductors.

\begin{acknowledgments}
The work is facilitated by the computer clusters at the Graphene Research Center at the National University of Singapore.
\end{acknowledgments}

\appendix

\section{Spectrum of density fluctuations for thin flake MoS$_2$ and TiSe$_2$}

\begin{figure*}[th]
\rotatebox[origin=c]{0}{\includegraphics[width=1.6\columnwidth]{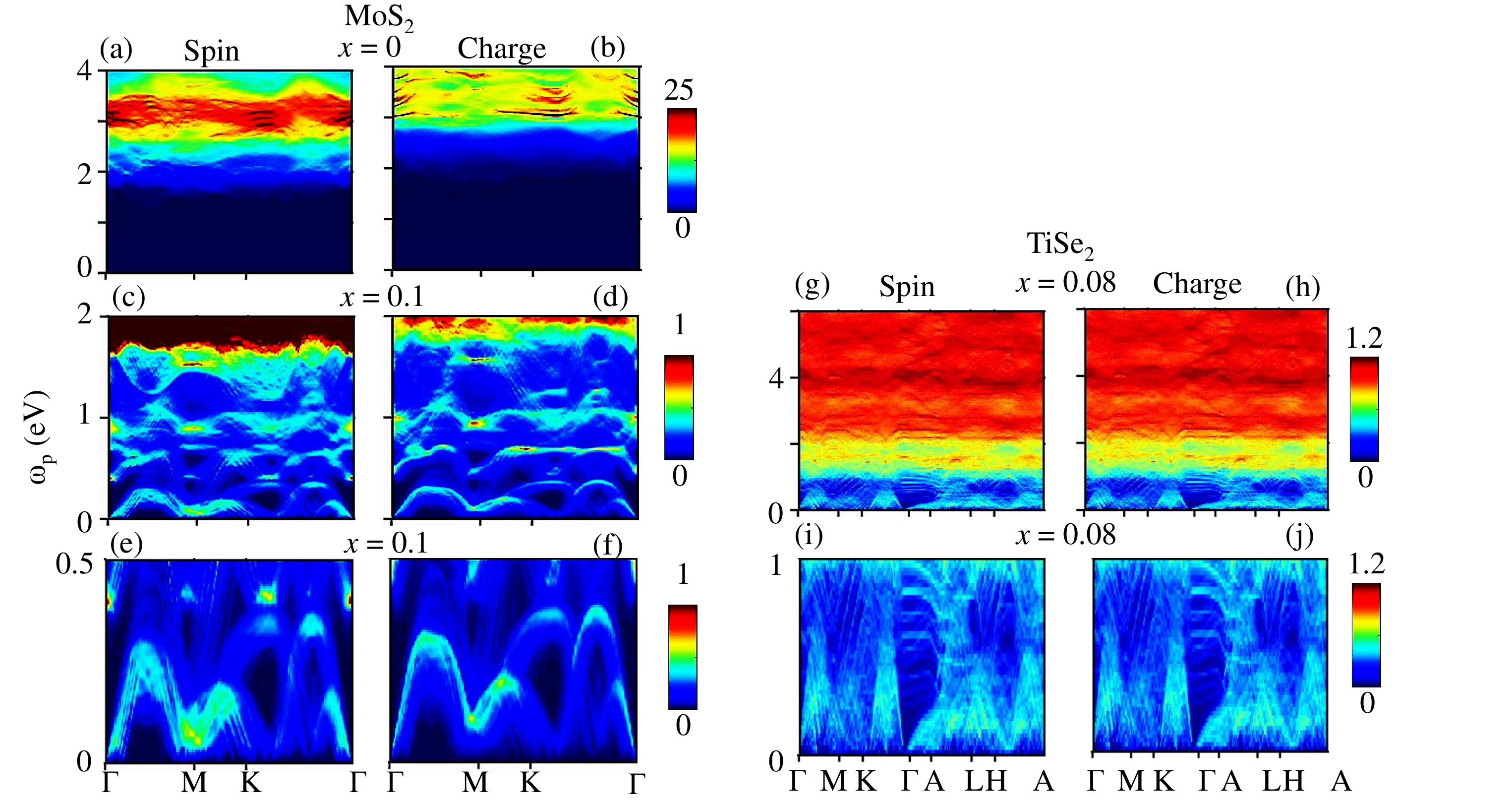}}
\caption{Imaginary parts of the spin and charge susceptibilities for various representative cases. (a)-(b) MoS$_2$ at $x=0$. (c)-(d), Same as (a-(b), but in at the optimal doping. (e)-(f) Zoomed in view of the same data shown in (c), and (d), respectively. (g)-(h), Same data, but for TiSe$_2$ at its optimal doping. (i)-(j), Zoomed in view of their corresponding upper panels. The results for TiSe$_2$ are plotted in the log-scale for clarity, while that for MoS$_2$ are plotted in linear scale. All results are shown by summing over all band contributions. 
}\label{fig5}
\end{figure*}

In the band insulating state of MoS$_2$, both spin and charge susceptibilities, shown in Figs.~\ref{fig5}(a-b), are gapped inside the particle-hole continuum. No collective excitation develops inside the spin gap in a band insulator, unlike in a magnetic insulators where a spin-wave dispersion (Goldstone mode) arises inside the particle-hole continuum due to symmetry breaking (see e.g., Fig.~6 of Ref.~\onlinecite{DasINS}). At finite doping (Figs.~\ref{fig5}(c-d)), in the metallic state, various intra-band transitions turn on and give rise to low-energy modes. Some of the modes extends to $\omega_p\rightarrow 0$. The spin channels are moved to lower in energy than the charge counterpart due to the many body correction within the RPA model. These low-energy modes are responsible for the increase of  renormalization effect at finite doping. In TiSe$_2$ at its optimal doping, due to the presence of many bands in the low-energy spectrum, the density fluctuation channels exhibit several modes, as shown in Figs.~\ref{fig5}(g-h). 

\section{Momentum and doping dependent self-energy}

\begin{figure}
\rotatebox[origin=c]{0}{\includegraphics[width=1.0\columnwidth]{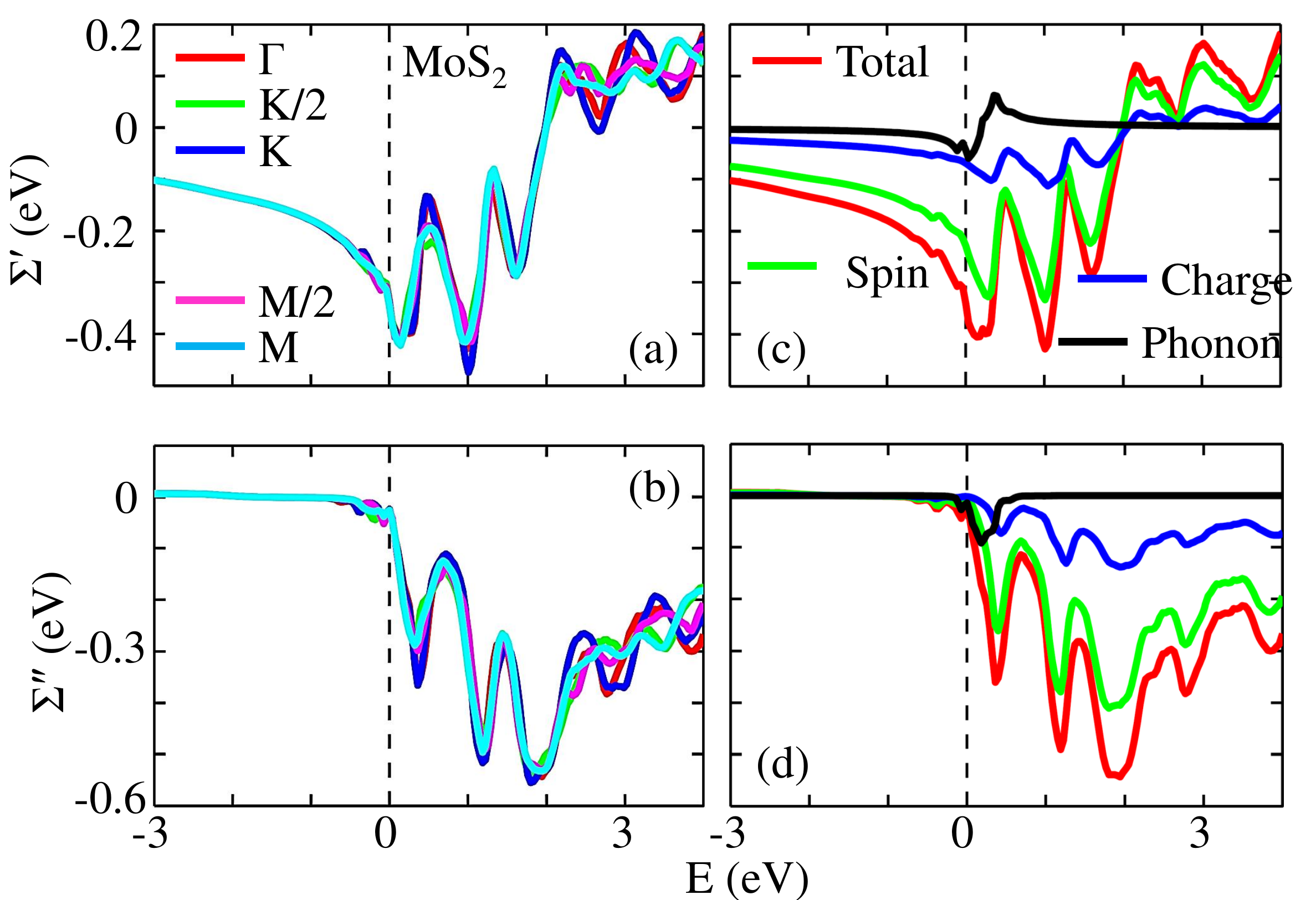}}
\caption{Analysis of various self-energy properties for MoS$_2$. (a)-(b) Momentum dependence of the real and imaginary parts of total self-energy at optimal doping for the band which crosses the Fermi level. (c)-(d) Various components of the total self-energy at the same doping. As mentioned in the main text, the largest contribution stems from the spin channel.  
}\label{fig6}
\end{figure}

\begin{figure}
\rotatebox[origin=c]{0}{\includegraphics[width=1.0\columnwidth]{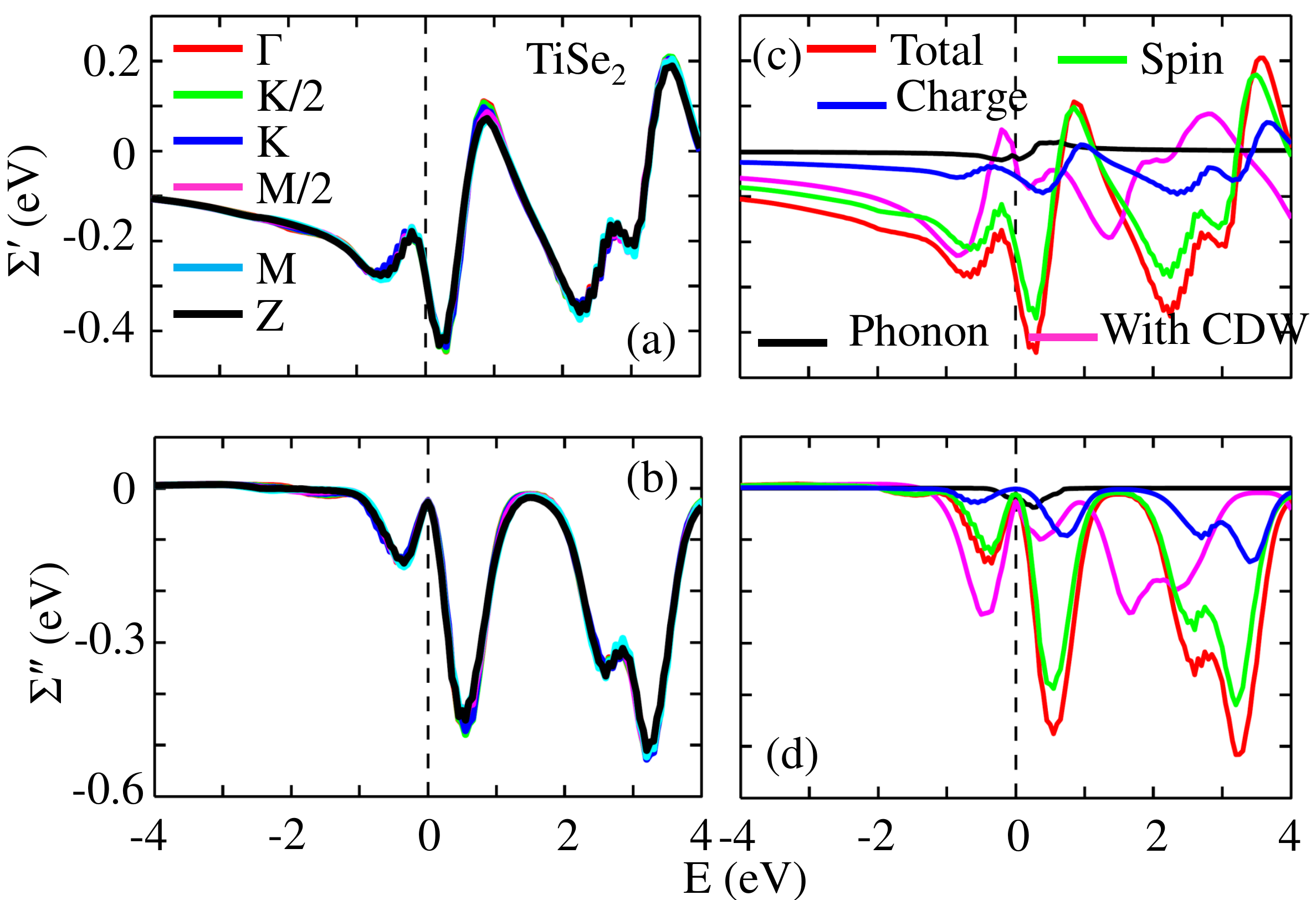}}
\caption{(a)-(b) The momentum dependence of the self-energy is reduced in this system than in MoS$_2$. The three-dimensionality of the self-energy is also weak. (c)-(d) As in MoS$_2$, the spin-density fluctuation is also dominant in TiSe$_2$. The CDW results are shown for $x=0.05$.  
}\label{fig7}
\end{figure}

The momentum dependence of the self-energies is one of the advantage of the present MRDF method, which allows us the understand the origin of the strong correlation feature from the band structure properties. As mentioned before, the density-density fluctuations arise from the JDOS fluctuation, which means it is dominated by the higher DOS regions in both filled and empty states. As seen in Fig.~\ref{fig6}, for MoS$_2$, there is a strong spectral weight of the spin and change mode around ${\bf q}=$ M point, which disperses strongly along the $\Gamma$ point, than along the K point. This low-energy mode is mainly responsible for the low-energy band renormalization in the self-energy, see Fig.~\ref{fig7}. In both real and imaginary part of the self-energy, the strongest renomalization appears around ${\bf k}$=K point, where two bands almost overlap, see Fig.~\ref{fig2}, giving rise to higher DOS at the Fermi level. In the right hand panel of Fig.~\ref{fig6}, it is shown that the spin contribution to the fluctuation spectrum is dominant, and then the charge contribution, while phonon contribution is the least. 

The momentum dependence of the self-energy is much weak in TiSe$_2$, as shown in Fig.~\ref{fig7}. This is also reflected in the the susceptibility spectrum in Fig.~\ref{fig5} which is quite broad and dispersive. Similarly, the electronic dispersion of this system also indicates that there is not much of an electronic `hot-spot' for the strong correlation phenomena.  Despite {\tise} having a metallic DFT band structure, its self-energy spectrum exhibits particle-hole asymmetry. This is due to the fact the the bands are more dispersive below $E_F$ than above it, which makes the correlation effects to be larger in the empty states. 

\section{Doping dependent $m^*/m_b$, renormalized $\omega_{\rm log}$ and $\mu^*$}

\begin{figure}[h]
\rotatebox[origin=c]{0}{\includegraphics[width=1.0\columnwidth]{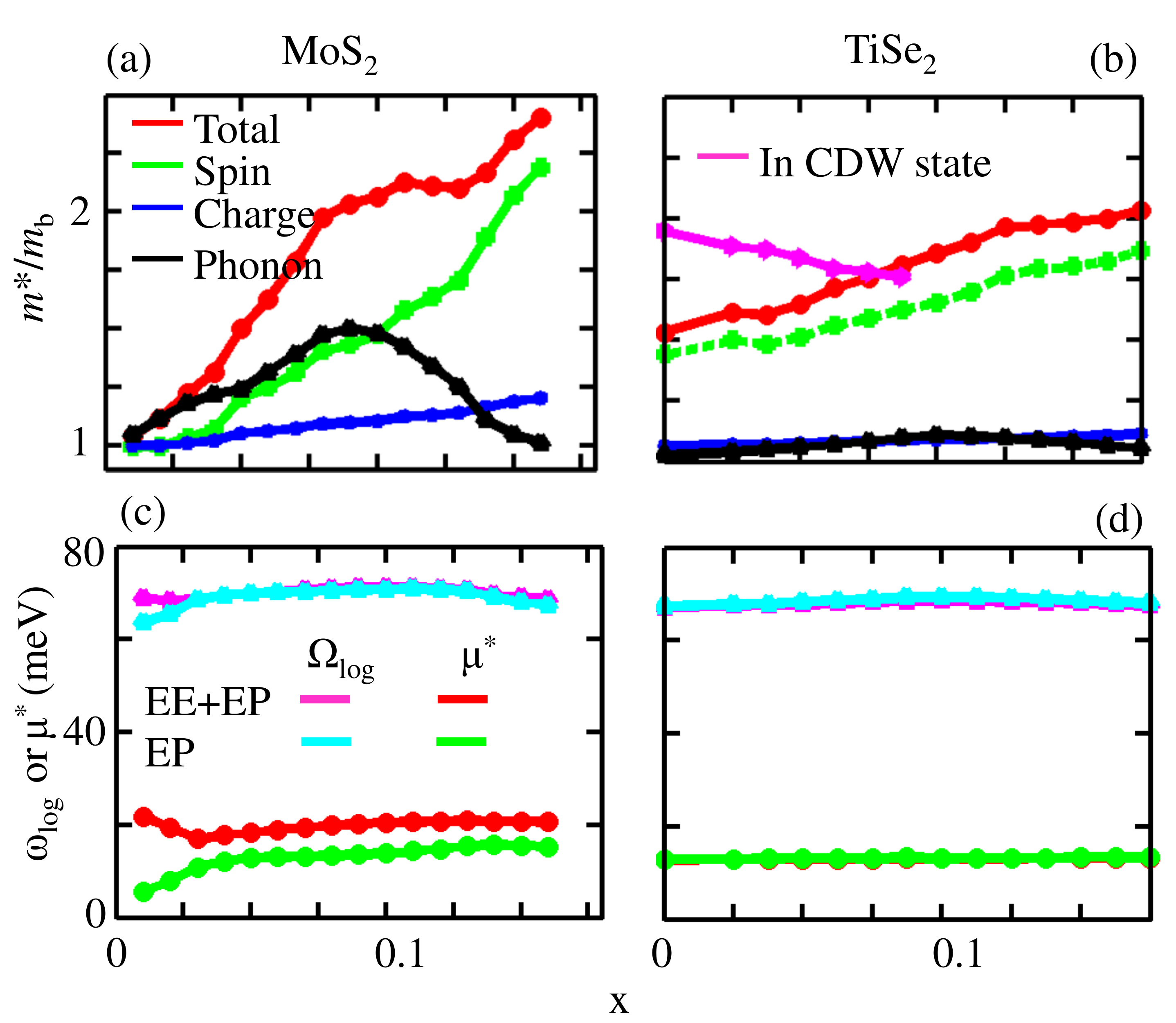}}
\caption{(a) The mass renormalization $m^*/m_b$ at the Fermi level for the highest renormalized band (nearest to the Fermi level) as a function of doping for MoS$_2$. (b) Same as (a), but for TiSe$_2$. (c) Doping dependence of renormalized $\omega_{\rm log}$ and $\mu^*$ for MoS$_2$. (d) Same as (c), but for TiSe$_2$. 
}\label{fig8}
\end{figure}

The mass enhancement $m^*/m_b=1/Z$ at the Fermi level is plotted in the upper panel of Fig.~\ref{fig8} for the highest renormalized band (nearest to the Fermi level) as a function of doping. As shown in Fig.~\ref{fig1}, the phonon part (black dots) exhibits a dome-like feature centering the optimal doping, while spin and charge contributions increase gradually with doping. As the phonon-contribution decreases in the overdoped region, the spin-contribution rises sharply. For TiSe$_2$, the phonon and charge contributions are very similar as a function of doping. In the CDW state, the renormalization is dominant in the undoped case, and then decreases monotonically with doping. All results are presented for one of the two bands in the low-energy spectrum in which the renormalization effect is strongest.  As deduced before, we get $m^*/m_b\sim$ 2 - 2.5 in the optimal to overdoped region in MoS$_2$, which is close to the experimental value of $\sim$2.4 for the hole band. Again for TiSe$_2$, the experimental value of $m^*/m_b\sim$1.74 is close to our result of 1.8 - 2. 

As demonstrated in the main text, the self-energy effects on the $\omega_{\rm log}$ and $\mu^*$ is rather small, and the values remain fairly doping independent, and do not exhibit any apparent anomaly at the optimal doping. More interestingly, the values are almost identical for EP and EE+EP cases, demonstrating that these features do not contribute to the formation of SC dome in these systems.


\begin{thebibliography}{99}
\bibitem{Note_Haas}In Ref.~\onlinecite{Haas}, the authors have found that $\omega_{\rm log}$ has an opposite doping dependent to EP coupling $\lambda$, which together creates a SC dome. After including EE interaction, we find that the doping dependence of the $\omega_{\rm log}$ is significantly reduced.
\end{thebibliography}
\end{document}